\documentclass[preprintnumbers,eqsecnum,amsmath,amssymb,twocolumn,nofootinbib]{revtex4}

\usepackage{graphicx}
\usepackage{dcolumn}
\usepackage{bm}
\usepackage{epsfig}

\newcommand{\be}{\begin{eqnarray}}
\newcommand{\ee}{\end{eqnarray}}

\newcommand{\ta}{\overline{\theta}}
\newcommand{\ds}[1]{#1{\hskip-2.2mm}/}
\newcommand{\dsl}[1]{#1{\hskip-1.7mm}/}

\begin{document}

\preprint{ECT*-04-05}
\preprint{RM3-TH/04-15}
\preprint{ROME1-1379/2004}

\vspace*{0.10truecm}

\title{The Neutron Electric Dipole Moment in the Instanton Vacuum:\\[2mm]
Quenched Versus Unquenched Simulations}

\author{P.~Faccioli$^a$, D.~Guadagnoli$^b$ and S.~Simula$^c$}

\affiliation{ \vspace{0.5cm}
$^a$E.C.T.$^\star$, Strada delle Tabarelle 286, I-38050 Trento,
    and INFN, Sezione Collegata di Trento, Trento, Italy\\
$^b$Dipartimento di Fisica, Universit\`a di Roma ``La Sapienza'', 
    and INFN, Sezione di Roma, P.le A.~Moro 2, I-00185 Rome, Italy\\
$^c$INFN, Sezione di Roma Tre, Via della Vasca Navale 84, I-00146, Roma, 
Italy}

\begin{abstract}

We investigate the role played by the fermionic determinant in the evaluation 
of the $CP$-violating neutron electric dipole moment (EDM) adopting the 
Instanton Liquid Model. Significant differences between quenched and unquenched
calculations are found. In the case of unquenched simulations the neutron 
EDM decreases linearly with the quark mass and is expected to vanish in the 
chiral limit. On the contrary, within the quenched approximation, the neutron
EDM increases as the quark mass decreases and is expected to diverge as $1/m^{N_f}$ in the 
chiral limit. We argue that such a qualitatively different behavior is a parameter-free, 
semi-classical prediction and occurs because the neutron EDM is 
sensitive to the topological structure of the vacuum. 
The present analysis suggests that quenched and unquenched lattice 
QCD simulations of the neutron  EDM as well as of other observables 
 governed by topology might show up important differences in the quark mass dependence,
for $m_q \lesssim \Lambda_{QCD}$. 

\end{abstract}

\maketitle

\section{\protect Introduction \label{sec:introduction}}

The electric dipole moment (EDM) of the neutron provides direct information on 
the violation of the parity and time-reversal symmetries. Both the strong and 
the electroweak sectors of the Standard Model (SM) can generate violations of 
the above symmetries. As for the strong sector, it is known \cite{tHooft} that 
a gauge-invariant definition of the QCD vacuum requires to supplement the 
classical action with an additional gauge-invariant and renormalizable term, 
which in Euclidean space reads
 \be
    \mathcal{S}_{QCD}^{\prime} & = & \mathcal{S}_{QCD} ~ + ~ \mathcal{S}_{\theta} ~ ,
    \label{L} \\
    \mathcal{S}_{\theta} & = & i ~ \theta ~ \frac 1{32 \pi^2} \int d^4x ~ 
    F_{\mu \nu} \widetilde{F}_{\mu \nu} ~ ,
    \label{Ltheta}
 \ee
where 
 \be
    Q = \frac 1{32 \pi^2} \int d^4x ~ F_{\mu \nu} \widetilde{F}_{\mu \nu}
    \label{Q}
  \ee
is the topological charge operator, $\widetilde{F}_{\mu \nu} = (1 / 2) ~ \varepsilon_{\mu 
\nu \alpha \beta} F_{\alpha \beta}$ is the dual gluon field strength 
(incorporating the strong coupling constant) and $\theta$ is a (real) 
dimensionless parameter. The term (\ref{Ltheta}) is a source of $CP$ violation 
and goes under the name of strong $\theta$-term.

A second independent physical origin for a term of the form (\ref{Ltheta}) 
comes from the weak sector of the SM. The observation of $CP$ violation 
in K-meson systems implies that the quark mass matrix $M$ is not real and 
the mass term in the Lagrangian has the general form
 \be
    \mathcal{L}_M = \overline{\psi }_i^R ~ M_{ij} ~ \psi^L + \overline{\psi}_i^L 
    ~ M_{ij}^{\dagger} ~ \psi ^R~.
 \ee
The mass matrix can be made real and diagonal by means of an appropriate
chiral rotation, which generates a shift in the $\theta$ parameter: 
 \be
    \theta \rightarrow \overline{\theta } \doteq \theta + \textrm{argdet}(M)~.
 \ee

The real constant $\ta$ is an additional dimensionless parameter, 
which has to be fixed from experiment. This can be done by exploiting the 
fact that (\ref{Ltheta}) is a source of $CP$ violation which leads to a 
non-vanishing value of the neutron EDM. At present there are only upper bounds 
on the neutron EDM and the most constraining one is $|\mathbf{d}_n| < 6.3 
\times 10^{-13} ~ (\mathrm{e \cdot fm})$ \cite{ILL}.

In order to translate this experimental information into a constraint on 
$\overline{\theta}$, one needs to compute the neutron EDM in QCD, including
the contribution of the topological term $i {\overline{\theta} ~ Q}$.
So far, this has been done only within model-dependent frameworks, starting 
from the works of Refs.~\cite{Baluni,Crewther}. In 
\cite{Baluni} Baluni performed a calculation of the EDM in the Bag Model 
and found $|\mathbf{d}_n| = 2.7 \times 10^{-3} ~ \overline{\theta } ~ 
(\mathrm{e \cdot fm})$. In \cite{Crewther} Crewther et al.~proposed an 
approach based on current algebra relations and found $|\mathbf{d}_n| = 3.6 
\times 10^{-3} ~\overline{\theta } ~ (\mathrm{e \cdot fm})$.

The above model calculations agree in pointing out that $|\overline{\theta}| 
\lesssim 2 \cdot 10^{-10}$. Understanding why $\ta$ is so small is an open 
challenge, which goes under the name of strong $CP$ problem. 

In order to estimate the relevant matrix element in a model-independent way,
a non-perturbative approach based on the fundamental theory, like lattice QCD,
is required. However a lattice estimate of the neutron EDM is not yet
available \cite{Aoki}. Recently \cite{Diego} a new strategy for computing
the neutron EDM in lattice QCD has been proposed. The starting point is the 
expansion of the matrix element of the EDM operator to lowest order in 
$\overline{\theta}$. This allows to remove the complex $\theta$-term 
(\ref{Ltheta}) from the action, but at the expense of calculating the matrix element 
of a composite operator made of the electromagnetic dipole and the topological 
charge operators:
 \be
    \mathbf{d}_n & = & \langle N(\ta)| \int d\mathbf{x\,\,x\,} J_0(\mathbf{x)} 
    |N(\ta) \rangle
    \label{Expandtheta} \\
    & \simeq & -i ~ \overline{\theta }\,\, \langle N(0)| \int d\mathbf{x\,\,x\,}
    J_0(\mathbf{x)} \nonumber \\
    & \cdot & \left[ \frac{1}{32 \pi ^2} \int d^4z ~ F_{\mu \nu } \widetilde{F}
    _{\mu \nu } \right] |N(0) \rangle
    \label{Dtopocharge}
 \ee
where $|N(\overline{\theta}) \rangle$ denotes the neutron state at a finite 
value of $\ta$.

Computing the matrix element (\ref{Dtopocharge}) on the lattice is 
a very challenging task. The main source of difficulties is that the topological 
charge operator is very noisy \cite{Gockeler}. A possible way-out, suggested in 
Refs.~\cite{Aoki,Diego}, consists in exploiting the anomalous Ward identities 
to replace the topological charge operator in (\ref{Dtopocharge}) with the 
pseudo-scalar density operator, namely 
 \be
    & & \langle \mathcal{O}\left[ \frac{2 N_f}{32 \pi ^2} \int d^4z ~ 
    F_{\mu \nu } \widetilde{F}_{\mu \nu }\right] \rangle = \nonumber \\ 
    & & - 2m_0 ~ \langle \mathcal{O} \int d^4z ~ P(z) \rangle_{disc.}~, 
    \label{Ward}
 \ee
where $\langle ~ \rangle $ denotes the quantum average over all configurations, 
$N_f$ is the number of quark flavors, $m_0$ is the bare quark mass ($ 1 / m_0 
\equiv (1 / m_u + 1 / m_d + 1 / m_s) / 3$ for $N_f = 3$) and $P$ is the
flavor-singlet pseudo-scalar operator $P(x) = \sum_f \overline{\psi}_f(x) 
\gamma _5 \psi_f(x)$. In the r.h.s.~of Eq.~(\ref{Ward}) $\langle ~ \rangle _{disc}$ 
denotes the quantum average obtained by retaining only the Wick contractions 
in which the pseudo-scalar operator is contracted in a virtual quark loop. 
The numerical applicability of such a strategy is still under investigation.

The use of the quenched approximation is quite natural for a first-time 
calculation. We point out however that neglecting the fermionic determinant may
result in a sizable systematic error in case of the neutron EDM, since the
evaluation of the latter involves flavor-singlet operators. Indeed, in case 
of hadron masses and decay constants quenched calculations turn out to be 
accurate, since they appear to agree with available (partially) unquenched 
simulations within $5 \div 10 \% $ accuracy \cite{CP-PACS,Davies}. However, in 
case of observables which are directly related to the topological properties 
of the vacuum, the contribution of the fermionic determinant may become more 
important. 

This can be seen, for example, by considering the Index Theorem, according to which 
the total topological charge of a gauge configuration relates to the difference 
of the number of left- and right-handed zero-modes of the Dirac operator:
 \be
    Q = n_L - n_R ~ .
    \label{index}
 \ee
On the other hand, the fermionic determinant can be written in terms of
eigenvalues of the Dirac operator as:
 \be
    \mathrm{det}(\ds{D} + m_f) = m^\nu \prod_{\lambda >0}(\lambda ^2+m_f^2)~,
    \label{product}
 \ee
where  $\ds{D} \,\psi _\lambda =\lambda \,\psi _\lambda $, and $\nu $ is the 
number of zero-modes.

Equations (\ref{index}) and (\ref{product}) imply that, for small values of the
bare quark masses, the fermionic determinant suppresses configurations with
a non-vanishing topological charge. Since the EDM is an observable which 
relates directly to topology, we argue that evaluating this quantity in 
quenched and unquenched lattice QCD might lead to quite different results. 
Of course, one can still hope that sufficiently away from the chiral limit 
quenched and full simulations give comparable results.

While waiting for lattice results on the neutron EDM, the aim of this work is 
to investigate the role played by the fermionic determinant using a model
which is expected to take into account properly topological effects. Such a
model is the Instanton Liquid Model (ILM), since instantons are 
topological gauge configurations which dominate the QCD Path Integral 
in the semi-classical limit. They generate the so-called 't~Hooft interaction,
that solves the U(1) problem \cite{tHooft} and spontaneously breaks chiral 
symmetry \cite{dyakonov}, but does not confine. Evidence for such an 
instanton-induced interaction in QCD comes from a number of  
phenomenological studies \cite{shuryakrev}, as well as from lattice 
simulations \cite{chu94,Degrand,scalar}. The ILM assumes that the QCD vacuum 
is saturated by an ensemble of instantons and anti-instantons. The only 
phenomenological parameters in the model are the instanton average 
size and density. Their values were estimated long ago from the global 
vacuum properties~\cite{shuryak82}. Since this model provides very successful 
descriptions of both the pion and nucleon electromagnetic structure 
\cite{nucleonFF} and of the topological properties of the QCD vacuum 
\cite{toposcreening}, we expect its prediction for the neutron EDM 
to be realistic.

The dynamical mechanism leading to the formation of the neutron EDM in the 
instanton vacuum was recently investigated by one of the author~\cite{bnumber}.
It was found that, during the tunneling processes, the $\theta$-term generates 
an effective repulsion between matter and anti-matter, in the neutron. As a 
consequence, quarks and  anti-quarks migrate in opposite directions. Hence, 
at least on the semi-classical level, the EDM arises from the local separation
of positive and negative baryonic charges in the neutron. It does not follow 
from the displacement of the positive and negative electric charge carried by 
the {\it valence} quarks, as one would intuitively expect in a naive 
non-relativistic quark model picture.

In this work we show that the use of the ILM clearly suggests that neglecting 
the fermionic determinant leads to a dangerous divergence in the chiral limit, 
which in the {\em full} model is regulated by topological screening. 
We will compare unquenched and quenched model calculations at values of the 
space-time volume and of the current quark mass which are comparable to 
those used in present-day lattice simulations, and provide numerical 
estimates of the neutron EDM.

In order to model quenched and full QCD, we will use the Interacting Instanton 
Liquid Model (IILM) developed by Shuryak and collaborators (see 
Ref.~\cite{shuryakrev} for a review). In this model the configurations of the 
instanton ensemble are generated by means of a Metropolis algorithm which
explicitly accounts for the fermionic determinant. The instanton size
distribution is not fixed a priori, but it is obtained dynamically. The
average density was fixed by minimizing the free energy~\cite{IILM}. 
In the IILM the neglect of the fermionic determinant generates pathologies 
which are analogous to the ones observed in quenched lattice QCD (see 
Ref.~\cite{scalar}).

The feasibility of instanton calculations relies on two important
simplifications, which arise when one restricts the functional integration
over the gauge field configurations to an ordinary integral over the
collective coordinates of an ensemble of $N_{+}$ instanton and $N_{-\;}$%
anti-instantons. On the one hand, one has to deal with a dramatically
smaller number of degrees of freedom, typically of the order of $12 \times
(N_{+}+N_{-}) \sim 10^3$ (for a typical box of the size of those used in
lattice simulations). On the other hand, evaluating the topological charge
operator in the instanton vacuum is trivial, because it amounts to simply
counting the number of instantons and anti-instantons in the ensemble 
(indeed $Q = N_{+} - N_{-}$ ). Due to these simplifications, the numerical 
simulations leading to the EDM can be performed on a regular workstation.

We show that, within the IILM, quenched simulations become unreliable for 
quark masses smaller than the strange quark mass. Indeed, for $m_q \gtrsim 
200$ MeV, quenched and full simulations give results which agree within 
statistical errors, whereas for quark masses of the order of $100$ MeV, 
quenched calculations overestimate the EDM by a factor $\simeq 4$. This is due 
to the appearance of the chiral divergence mentioned previously. Moreover, we 
stress that other quantities, like e.g.~the nucleon mass or the quark 
condensate, are not drastically affected by quenching.

Assuming a linear mass dependence, the extrapolation of the unquenched ILM 
results to a physical value of the light quark mass between 4 and 10 MeV 
yields: $|{\bf d}_n| = (6 \div 14) \times 10^{-3}~\ta$~(e~$\cdot$~fm), which 
is a factor about $2 \div 4$ larger than the estimates obtained in 
Refs.~\cite{Baluni,Crewther}.

The paper is organized as follows. In Section \ref{sec:correlations} we 
shall discuss our method for relating the neutron EDM to Euclidean correlation
functions. In Section \ref{sec:EDM} we will study these correlators in the
instanton vacuum and derive the quenched and unquenched expressions for the 
neutron EDM. In Section \ref{sec:chiral} we analyze the chiral behavior 
both for quenched and unquenched calculations. Results are presented and 
discussed in Section \ref{sec:results}, and, finally, conclusions are 
summarized in Section \ref{sec:conclusions}.

\section{Relating the neutron EDM to euclidean correlation functions 
\label{sec:correlations}}

To compute the neutron EDM in a field-theory framework, we start by 
considering the following Euclidean correlation function:
 \be
    D_{\tau}^{\gamma \gamma^{\prime }}(\mathbf{p,q)} &\doteq &\int 
    d\mathbf{x}\int d\mathbf{y} e^{i \mathbf{p x} + i\mathbf{q y}} y_3 
    G_{\tau}^{\gamma \gamma^{\prime }}(\mathbf{x,y)}
    \label{D3def} \\
    &=& -i \frac{\partial}{\partial q_3} \int d\mathbf{x} \int d\mathbf{y} 
    e^{i \mathbf{p x} + i \mathbf{q y}} G_{\tau}^{\gamma \gamma^{\prime }}
    (\mathbf{x, y}) ~ , ~~~~ \nonumber \\
    G_{\tau }^{\gamma \gamma^{\prime }} (\mathbf{x,y)} &\doteq & \langle
    0| T [J_N^\gamma (\mathbf{0},2\tau) J_4(\mathbf{y},\tau) \overline{J}
    _N^{\gamma^{\prime}} (\mathbf{x},0) |0\rangle
    \label{G3def}
 \ee
where $J_N^\gamma(x) = \varepsilon_{a b c}(u^T_a C \gamma_5 d_b) d_c^\gamma$ 
is an interpolating operator for the neutron, $p = (\mathbf{p},\omega)$, 
$p^{\prime} = (\mathbf{p} + \mathbf{q}, \omega^{\prime })$. In the limit 
of large Euclidean times $\tau $ one can show that: 
 \be
    D_{\tau}^{\gamma \gamma^{\prime}}(\mathbf{p,q)} 
    & \stackrel{\tau \rightarrow \infty}{\longrightarrow} & 
    -i \frac{\partial}{\partial q_3} \left( Z^2 \frac{1}{2\omega ~ 
    2\omega^{\prime}} e^{-(\omega^{\prime} + \omega) \tau} \right. 
    \nonumber \\
    & \cdot & \left. \left[ (\dsl{p}^{\prime} + M) ~ \mathrm{K}(q^2)
    (\dsl{p} + M) \right]^{\gamma \gamma^{\prime}} 
    \right) ~ , ~~~~ 
\ee
where $M$ is the mass of the neutron, $Z$ is a constant defined by
 \be
    \langle 0 | J_N(0) | N(p, s) \rangle = Z ~ u(p, s) ~ ,
    \label{Z}
 \ee
where $u(p, s)$ is a Dirac spinor, and 
 \be
    \mathrm{K}(q^2) & = & F_1(q^2) ~ \gamma_4 + i F_2(q^2) ~
    \frac{\sigma_{4 3}}{2M} ~ q_3 \nonumber \\
    & + & F_A(q^2) ~ (q^2 \gamma_4 -2M q_4) ~ \gamma_5 
    \nonumber \\
    & + & F_3(q^2) ~ \frac{\sigma_{43}}{2M} ~ \gamma_5 ~ q_3 ~ .
 \ee
with $\sigma_{43}$ the Euclidean version of $\sigma_{03}=i\gamma_0 \gamma_3$.
Notice that the form factors $F_A(q^2)$ and $F_3(q^2)$ do not appear if we
assume that the matrix element is invariant under $C$, $P$, and $T$. In
particular $F_3(q^2 = 0)$ relates directly to the EDM through the relationship:
 \be
    D \doteq \frac{|\mathbf{d}_n|}e = \frac{F_3(0)}{2M} ~ .
 \ee
In order to isolate the contribution to the EDM it is convenient to define
the following trace:
 \be
    \Sigma_\tau (\mathbf{p}, \mathbf{q}) \doteq \mathrm{Tr}[D_\tau 
    (\mathbf{p}, \mathbf{q}) ~ \Gamma_3], \qquad \Gamma_3 \doteq
    \left( \begin{array}{ll} \sigma _3 & 0 \\ 
     0 & \sigma_3 \end{array} \right) 
 \ee
and obtain: 
 \be
    \Sigma_\tau (\mathbf{p},\mathbf{q}) & \longrightarrow & -\frac{Z^2}
    {8M} i \frac{\partial}{\partial q_3} \Phi_\tau (\mathbf{p},\mathbf{q}) , \\
    \Phi_\tau (\mathbf{p},\mathbf{q}) & = & \frac{1}{\omega \omega^{\prime}}
    ~ e^{-(\omega +\omega^{\prime}) \tau} \mathrm{Tr}[(\dsl{p}^{\prime} + M) ~ 
    \gamma_5 ~ \sigma_{43} \nonumber \\
    & \cdot & (\dsl{p} + M) \Gamma_3] q_3 ~ F_3(q^2) ~ .
 \ee
Taking $\mathbf{q} = 0$ we obtain:
 \be
    \Sigma_\tau (\mathbf{p,q = 0}) 
    \stackrel{\tau \rightarrow \infty}{\longrightarrow}
     -2 Z^2 e^{-2 \omega \tau} 
    \frac{p_1^2 + p_2^2}{\mathbf{p}^2 + M^2} ~ D ~ .
 \ee

\begin{figure}

\includegraphics[scale=0.5,clip=]{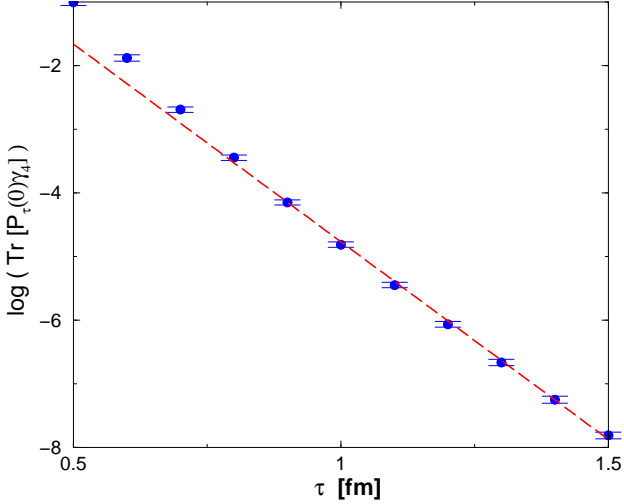}

\caption{ILM results for $log( Tr [P_\tau ({\bf p}={\bf 0})~\gamma_4] )$. The 
linear behavior denotes the isolation of the nucleon pole. The slope of the 
line corresponds to the nucleon mass $M$ (in units of inverse fm) while
its intercept corresponds to $log(2\,Z^2)$. These results were obtained in 
the quenched approximation, by averaging over the configurations of an 
ensemble of $56$ pseudo-particles in a volume $V = 2.4^3 \times 4 = 56$ fm$^4$ 
and quark masses of $m_q = 1.00 ~ \Lambda_{QCD}$.}

\label{fig:twopoint}

\end{figure}

All correlation functions defined so far are projected onto fixed momentum
components, hence the spatial position of the sources and of the
electromagnetic operator are completely unspecified. From the practical
point of view, evaluating these quantities can be very challenging, because
they involve a six-dimensional integration over the two spatial
hyperplanes. This has to be performed numerically and requires to compute a
large number of evaluation points for the integrand, each of which is a
three-point correlation function. However, as long as we are interested in a 
static property of the neutron, such as the EDM, it is possible to
reduce the number of spatial integrations by working in a mixed
space-momentum representation, where one projects only on the momentum
transfer by the photon and leaves the source and the sink at a fixed spatial
location. In this way one gives up the information about the initial momentum
of the neutron. Let us define the correlator:
 \be
    \widetilde{\Sigma }_\tau (\mathbf{x},\mathbf{q} = \mathbf{0) = } \int 
    \frac{d\mathbf{p}}{(2 \pi)^3} ~ e^{-i\mathbf{p\,x}} 
    \Sigma_\tau(\mathbf{p}, \mathbf{q} = \mathbf{0}) ~ .
 \ee
Choosing $\mathbf{x} = \mathbf{0}$ we obtain: 
 \be
    \widetilde{\Sigma }_\tau (\mathbf{x} = \mathbf{0},\mathbf{q} = 
    \mathbf{0)} = \int d\mathbf{y} ~ y_3 ~ \mathrm{Tr} \mathbf{[} 
    G_\tau(\mathbf{x} = \mathbf{0}, \mathbf{y}) ~ \Gamma _3] 
    \nonumber \\
    \stackrel{\tau \rightarrow \infty}{\longrightarrow}  D \times
    \left( -2 Z^2 \int \frac{d\mathbf{p}}{(2\pi)^3} ~ e^{-2 \omega \tau} 
    ~ \frac{p_1^2 + p_2^2}{\mathbf{p}^2 + M^2}~\right)~.~
    \label{sigma}
 \ee

Evaluating $\widetilde{\Sigma }_\tau (\mathbf{x} = \mathbf{0},\mathbf{q} =
\mathbf{0})$ involves only a three-dimensional integration over the position
of the electromagnetic current. Moreover, the number of independent
numerical integrations can be further reduced to two, by exploiting the
rotational symmetry of $G_\tau (\mathbf{x} = \mathbf{0},\mathbf{y})$ around
the $y_3$ axis:
 \be
    \widetilde{\Sigma }_\tau (\mathbf{x} = \mathbf{0}, \mathbf{q} = \mathbf{0}) 
    & = & 2\pi \int_{-\infty }^\infty dy_3 ~ y_3 ~ \int d|\mathbf{y}_{\perp}|
     ~ |\mathbf{y}_{\perp}| \nonumber \\
    & \cdot & \mathrm{Tr}[G_\tau(\mathbf{x} = \mathbf{0}, \mathbf{y}) ~ 
    \Gamma_3] ~ .
    \label{sigmatilde}
 \ee

As usual, the neutron mass $M$ and the coupling $Z$ to the neutron 
interpolating operator appearing in the r.h.s.~of Eq.~(\ref{sigma}) 
can be extracted from the zero 
momentum two-point correlator, 
 \be
    P_\tau^{\gamma \gamma^{\prime}}(\mathbf{p} = \mathbf{0}) & = & \int 
    d\mathbf{x} \langle 0| T [J_N^\gamma(\mathbf{x},\tau) 
    \overline{J}_N^{\gamma^{\prime }}(\mathbf{0}, 0)| 0\rangle 
    \nonumber \\
    & \stackrel{\tau \rightarrow \infty}{\longrightarrow} & 
    Z^2 \frac{(1 + \gamma_4)^{\gamma \gamma^{\prime }}} 2 e^{-M \tau} ~ .
 \ee

\section{The neutron EDM in the instanton vacuum \label{sec:EDM}}

In QCD the neutron EDM is non-zero only at finite values of the 
$\overline{\theta}$-angle. To lowest order in this parameter 
the three-point correlation function (\ref{G3def}) reads:
 \be
    G_\tau^{\gamma \gamma^{\prime}}(\mathbf{x,y)} & = & i \overline{\theta}
    \langle 0| T [J_N^\gamma (\mathbf{0}, 2\tau ) 
    J_4(\mathbf{y},\tau) \overline{J}_N^{\gamma^{\prime }}(\mathbf{x},0)  
    \nonumber \\
    & \cdot & \frac{1}{32 \pi^2}\int d^4z ~ F_{\mu \nu}(z) ~ \widetilde{F}_{\mu 
    \nu}(z)] |0 \rangle ~ , ~~~~
    \label{EDMtheta}
 \ee
where $| 0 \rangle$ refers to the vacuum at $\overline{\theta} = 0$.

In the instanton vacuum the topological charge is condensed around
instantons and anti-instantons, and correlation functions are computed by
averaging over the configurations of a grand-canonical ensemble of such
pseudo-particles with a partition function given by: 
 \be
    \mathcal{Z}(\mu ,\overline{\theta }) & = & \sum_{N_{+},N_{-}} e^{(\mu +i
    \overline{\theta }) ~ N_{+} + (\mu - i\overline{\theta }) ~ N_{-}}
    ~ Z_{N_{+} N_{-}}  \label{Partition1} \\
    Z_{N_{+} N_{-}} & = &\frac{1}{N_{+}! N_{-}!} \prod_{i=1}^{N_{+} + N_{-}}
    \int d^4z_i d\rho _i dU_i d(\rho _i) \nonumber \\
    & \cdot & \prod_f\mathrm{det(}\ds{D} + m_f\mathrm{)\,} ~e^{-S_{int}} ~ .
 \ee
In this formula $S_{int}$ is the bosonic instanton-instanton interaction, 
$z_i$ denotes the position of the pseudo-particle of size $\rho _i$, $dU_i$ 
is the Haar measure normalized to unity and $d(\rho _i)$ is the instanton size 
distribution. For small-sized instantons, such a distribution can be calculated 
by integrating over small gaussian quantum fluctuations around the instanton 
solution. At two loops, 't Hooft originally found: 
 \be
    d(\rho) & = & \frac{0.466 e^{-1.679 N_c}}{(N_c-1)!(N_c-2)!} \left( 
    \frac{8 \pi^2}{g^2(\rho)} \right)^{2N_c} \nonumber \\
    & \cdot & \exp\left(-\frac{8 \pi^2}{g^2(\rho)}\right) \frac{1}
    {\rho^5} ~ .
 \ee
As the size of the instanton increases, the overlapping with the field of
other pseudo-particles becomes not negligible. This generates interaction
between pseudo-particles which dynamically suppresses the instanton weight
$d(\rho)$ for large $\rho$. From variational calculations \cite{dia}, 
phenomenological estimates \cite{shuryak82} and lattice simulations 
\cite{lattice} one finds that the $d(\rho)$ is peaked around a mean value 
$\overline{\rho} = 0.3 \div 0.5$ fm.

In (\ref{Partition1}) we have introduced a complex chemical potential 
$(\mu \pm i\overline{\theta}),$ associated with the fluctuations
of the number of instantons and anti-instantons in the ensemble. Equivalently,
one can rewrite the partition function in (\ref{Partition1}) as a sum over
the total topological charge and total number of pseudo-particles:
 \be
    \mathcal{Z}(\mu, \overline{\theta}) & = & \sum_{Q =0,\pm1,...}
    \sum_{N=1,2,...} e^{\mu N + i\overline{\theta } Q } Z_{Q N} \\
    Q & \doteq & N_{+} - N_{-} ~ , \\
    N & \doteq & N_{+} + N_{-} ~ .
 \ee
From this equation it follows that $i\overline{\theta }$ can be interpreted
as an imaginary chemical potential associated to the fluctuations of the
topological charge.

The relative width of the fluctuations of $N$ and $Q$ around their mean 
values is proportional to the inverse of the size of the system. Thus, in 
the thermodynamic limit, such fluctuations have no consequences on quantities 
that are intensive in $N$ and $Q$. Hence observables such as the nucleon mass 
can be reliably calculated assuming $N = \langle N \rangle$ and $Q = \langle Q 
\rangle = 0$. On the contrary fluctuations can never be neglected when 
computing averages of operators which are extensive in $N$ or $Q$, like 
the case of the correlation function (\ref{EDMtheta}).

Since in the ILM topology is clustered around instantons and anti-instantons, 
a generic matrix element of the type 
 \be
    \langle 0| \mathcal{O} \frac{1}{32 \pi^2} \int d^4z F_{\mu \nu }
    \widetilde{F}_{\mu \nu } |0 \rangle
 \ee
can be written as \cite{DPW96}
 \be
    \left \langle \mathcal{O} \frac{1}{32 \pi^2} \int d^4z F_{\mu \nu} 
    \widetilde{F}_{\mu \nu }\right\rangle = \sum_Q \mathcal{P}(Q ) ~ Q
    \left \langle \mathcal{O} \right \rangle_Q ~ ,
    \label{topoweight}
 \ee
where $\mathcal{P}(Q )$ denotes the relative occurrence of configurations 
with topological charge $Q$ and $\langle \mathcal{O} \rangle_Q$ is the
expectation value of the operator $\mathcal{O}$ at a fixed topological charge 
$Q$. Using the fact that $\sqrt{Q^2} / N \rightarrow 0$ for large $N$, Diakonov, 
Polyakov and Weiss obtained the following factorized result:
 \be
    \left\langle \mathcal{O} \frac{1}{32 \pi^2} \int d^4z F_{\mu \nu}
    \widetilde{F}_{\mu \nu} \right\rangle = \langle Q^2 \rangle \left( 
    \frac{d}{dQ} \left\langle \mathcal{O} \right\rangle \right)_{Q =0} ~ ,~~
    \label{chiral}
 \ee
where $\langle Q^2 \rangle $ is the topological susceptibility, which in full 
QCD and for small quark masses can be expressed in terms of the quark 
condensate, viz.
 \be
    \langle Q^2 \rangle = -\frac{V \langle \overline{\psi } \psi \rangle}
    {\sum_f m_f^{-1}} ~ .
    \label{topos}
 \ee
This relation holds as well in the instanton vacuum, if the fermionic 
determinant is taken into account~\cite{DPW96}, whereas in the quenched 
approximation one has 
 \be
    \langle Q^2 \rangle = N_+ + N_- = N ~ .
    \label{topos_quenched}
 \ee

Using (\ref{chiral}) and (\ref{topos}) the neutron EDM in the unquenched case
can be expressed as a function of a ratio of correlation functions and of
the $\overline{\theta}$-angle parameter:
 \be
    D^{(unq.)} = \overline{\theta} \frac{V |\langle \overline{\psi} \psi \rangle|}
    {\sum_f m_f^{-1}} \frac{\left|\frac{d}{dQ} \left( \widetilde{\Sigma}_\tau 
    (\mathbf{0},\mathbf{0)} \right)_{Q = 0}\right|}{ 2 Z^2 \int \frac{
    d\mathbf{p}}{(2 \pi)^3} ~ e^{-2 \omega \tau} \frac{p_1^2 + p_2^2}
    {\mathbf{p}^2 + M^2}}~, ~~~~
    \label{full}
 \ee
where the quark condensate, the coupling and the mass can be computed in the
IILM from the appropriate correlation functions.

Similarly, in the quenched approximation the electric dipole moment reads
 \be
    D^{(q.)} = \overline{\theta} \frac{N \left|\frac{d}{dQ} \left( 
    \widetilde{\Sigma}_\tau (\mathbf{0},\mathbf{0)} \right)_{Q = 0}\right|}{2 
    Z^2 \int \frac{d\mathbf{p}}{(2 \pi)^3} e^{-2 \omega \tau} 
    \frac{p_1^2 + p_2^2}{\mathbf{p}^2 + M^2}}~.
    \label{quenched}
 \ee

Notice that the above Eqs.~(\ref{full}) and (\ref{quenched}) are well defined 
in the thermodynamic limit, because the derivative of $\widetilde{\Sigma}_\tau$ 
is proportional to the inverse volume and the number of pseudo-particles $N$ 
grows linearly with the volume.

\begin{figure}

\includegraphics[scale=0.5,clip=]{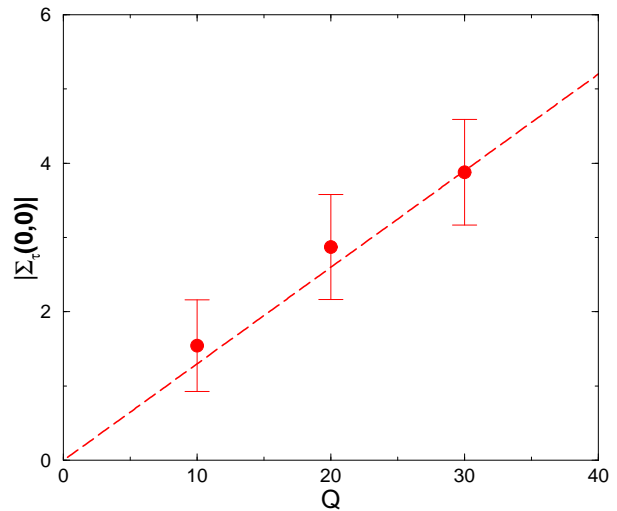}

\caption{Absolute value of the correlation function 
$\widetilde{\Sigma}_\tau({\bf 0},{\bf 0})$ as a function of the topological 
charge $Q$. Notice that the correlator (\ref{sigmatilde}), 
which relates directly to the EDM, receives contributions only 
from topologically non-trivial sectors, as expected.
These results were obtained in the quenched approximation, 
by averaging over the configurations of an ensemble of $130$ 
pseudo-particles in a volume $V = 3.1^3 \times 4 = 120$ fm$^4$ and quark 
masses of $m_q = 0.65 ~ \Lambda_{QCD}$.}

\label{fig:slope}

\end{figure}

\section{Chiral behavior of the neutron EDM within the IILM \label{sec:chiral}}

In this Section we derive the qualitative behavior of the expressions 
(\ref{full}) and (\ref{quenched}) in the chiral limit by studying the two 
quantities appearing in the r.h.s.~of Eq.~(\ref{chiral}). We will not consider 
corrections arising from chiral loops.

Let us start with the unquenched case, where the topological susceptibility 
$\langle Q^2 \rangle$ vanishes linearly in the quark mass
 \be
    \langle Q^2 \rangle \propto m ~ .
 \ee
Let us now look at the second factor in the r.h.s.~of Eq.~(\ref{chiral}). 
Writing out the discrete derivative explicitly one has
 \be
   \left( \frac{d}{dQ} \langle \mathcal{O} \rangle \right)_{Q = 0} = 
   \frac{\langle \mathcal{O} \rangle_{Q = 1} - 
   \langle \mathcal{O} \rangle_{Q = 0}}{(Q =1)} = 
   \langle \mathcal{O} \rangle_{Q =1} ~ ,
 \ee
where we have used the fact that the average of our $CP$ violating operator
vanishes in trivial topological sectors ($Q = 0$). Writing the average 
$\langle \mathcal{O} \rangle_Q$ explicitly in terms of functional integrals one 
gets
 \be
    \langle O \rangle_Q & = & \frac{1}{\mathcal{Z}} \int_{(Q)} \mathcal{D} 
    A_\mu \prod_f \left[ m_f^Q \prod_{\lambda > 0} \left( \lambda^2 + m_f^2 \right)
    \right]  \nonumber \\
    &\cdot & e^{-S_{YM}} ~ \mathrm{Tr}[SS...] ~ ,
 \ee
where $\mathrm{Tr}[SS...]$ denotes a trace of propagators arising from the 
explicit integration over the fermionic fields (Wick contractions). 

\begin{figure}

\includegraphics[scale=0.5,clip=]{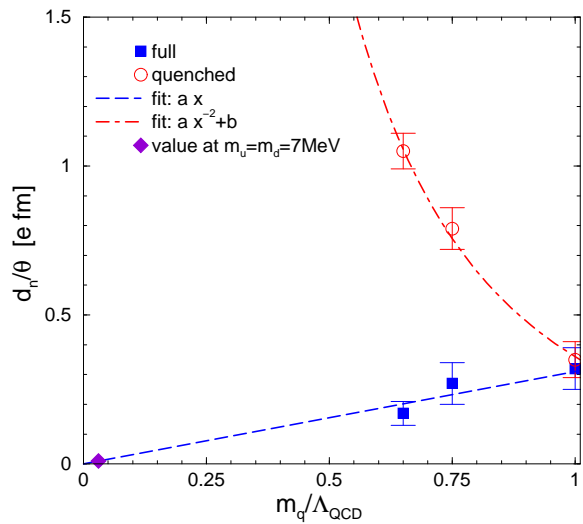}

\caption{ILM results obtained at different values of the quark masses with 
quenched (circles) and unquenched (squares) simulations. The behavior of the 
unquenched and quenched calculations for small quark masses is consistent with 
Eqs.~(\ref{Dfull_chiral}) and (\ref{Dq_chiral}), respectively.}

\label{fig:edm}

\end{figure}

In order to study the chiral behavior of the trace we recall that a quark
propagator in the given background can be written as: 
 \be
    S(x, y) & = & \sum_\lambda \frac{\psi_\lambda(x) \psi_\lambda^{\dagger}(y)}
    {\lambda - im_f}~, \\
    \ds{D} ~ \psi_\lambda(x)  & = & \lambda ~ \psi_\lambda(x)~.
 \ee

In a topologically non-trivial sector of the instanton vacuum, the Dirac
operator has $Q$ {\em exact} zero-modes associated to the {\em extra}
instantons (or anti-instantons) in the ensemble:
 \be
    \ds{D} ~ \psi_0^j = 0 ~ , \qquad j=1,...,Q~. 
 \ee
The quark propagator can then be written as:
 \be
    S(x, y) & = & \sum_{j=1}^Q \frac{\psi_0^j(x) \psi_0^{j \dagger}(y)}{-im_f}
    + \sum_{\lambda \neq 0}\frac{\psi_\lambda(x) \psi_\lambda ^{\dagger}(y)}
    {\lambda -im_f} \nonumber \\
    & \doteq & S_{top}(x, y) + S_{non-top}(x, y) ~ .
 \ee
In this expression $S_{top}(x, y)$ is the topological contribution to the
propagator arising from the exact zero-modes.

It is important to recall that the Pauli principle implies that 
at most $N_f$ quarks can propagate in the exact zero-mode wave function of an 
instanton. Thus one has
 \be
    \mathrm{Tr}[SS...] \propto \frac{1}{m^{N_f}} ~ .
 \ee

Collecting the results for chiral behavior of all the terms relevant for the EDM,
one gets
 \be
    D^{(unq.)} \propto \langle Q^2 \rangle \cdot \langle \mathcal{O} \rangle_{Q =1} 
    \propto m \cdot m^{N_f} \cdot \frac{1}{m^{N_f}} \propto m ~ .
    \label{Dfull_chiral}
 \ee

Hence we conclude that, when the fermionic determinant is included, the neutron 
EDM vanishes linearly in the quark mass, as it is also expected from other 
estimates based on chiral perturbation theory (see, e.g., Ref.~\cite{Crewther}).

When the quenched approximation is adopted, the topological susceptibility 
$\langle Q^2 \rangle$ becomes independent of the quark mass, while neglecting 
the fermionic determinant removes a factor $m^{N_f}$ from the numerator of the 
r.h.s.~of Eq.~(\ref{Dfull_chiral}). Therefore one expects a divergent behavior 
in the chiral limit of the form
 \be
    D^{(q.)} \propto \frac{1}{m^{N_f}} ~ .
    \label{Dq_chiral}
 \ee

We stress the fact the mass dependencies given in Eqs.~(\ref{Dfull_chiral}) 
and (\ref{Dq_chiral}) do not depend on the particular values of the model 
parameters which define the ILM. These results rely only on the working 
assumption that the quantum mixing  of the QCD vacuum can be described in 
terms of isolated tunneling events (instantons). In such a semi-classical 
limit, the Index Theorem is realized in a very specific way, with $N_f$ exact 
zero-modes associated to each unit of topological charge.

\section{Numerical results \label{sec:results}}

We have performed quenched and unquenched numerical simulations in the IILM, 
assuming two active degenerate flavors. We have used three values of the quark mass $m$, 
namely $m = 0.65, 0.75, 1.0$, in units of the Pauli-Villars scale
parameter $\Lambda_{QCD} = 220$ MeV~\cite{shuryakrev}. 
Correspondingly, we have used three periodic boxes of volume given by  
$V= (3.1^3\times 4),~ (2.8^3\times 4)$ and $(2.4^3\times 4)$~fm$^4$.
The total density of pseudo-particles in these ensembles 
has been chosen to satisfy $(N_+ + N_-) / V = 1$~fm$^{-4}$.

We have computed the quantities which are needed to determine $D / 
\overline{\theta}$ through Eqs.~(\ref{full}) and (\ref{quenched}).
The neutron mass $M$ and the coupling $Z$ of the interpolating operator 
have been obtained by extracting the slope and intercept of the logarithm 
of the two-point function  (see the example in Fig.~\ref{fig:twopoint}). 
The derivative of (\ref{sigmatilde}) with respect to $Q$  has been
 computed by varying by a small amount the 
number of instantons relative to that of anti-instantons in the box, while keeping the 
total number of pseudo-particles fixed ~(see the example in Fig.~\ref{fig:slope}). 

The results of our calculations are summarized in Tables~\ref{tab:tab1}-\ref{tab:tab2}. 
From Table~\ref{tab:tab1} we observe that, in case of the quark condensate, the neutron 
mass and the constant $Z$, quenched and unquenched simulations give almost the same 
results. This is as expected, since these quantities are not directly related the topological structure
of the QCD vacuum. On the contrary, the quantities reported in Table~\ref{tab:tab2} show a much 
larger sensitivity to the effects of the inclusion of the fermionic determinant. In 
particular the topological susceptibility $\langle Q^2 \rangle$ [see Eqs.~( \ref{topos}) 
and (\ref{topos_quenched})] and the EDM $D / \overline{\theta}$ [see Eqs.~(\ref{full}) 
and (\ref{quenched})] can differ by a factor of $\simeq 3 \div 4$ between the quenched 
and unquenched simulations with our choice of the quark mass values.

\begin{table}

\begin{center}

\begin{tabular}{|c|c|c|c|} \hline

$\stackrel{\mbox{V = 56 fm}^4}{\mbox{m = 1.00 }\Lambda_{QCD}}$
& $|\mathbf{\langle} \overline{\psi} \psi \mathbf{\rangle}|^{1/3}$ [GeV] 
& $M_N$ [GeV] & $Z$ [GeV$^6$]
\\ \hline
quenched & 0.186 & 1.42 & 0.018 
\\ \hline
unquenched & 0.186 & 1.41 & 0.017
\\ \hline
\end{tabular}

\vspace{0.20cm}

\begin{tabular}{|c|c|c|c|} \hline

$\stackrel{\mbox{V = 90 fm}^4}{\mbox{m = 0.75 }\Lambda_{QCD}}$
& $|\mathbf{\langle} \overline{\psi} \psi \mathbf{\rangle}|^{1/3}$ [GeV] 
& $M_N$ [GeV] & $Z$ [GeV$^6$]
\\ \hline
quenched & 0.199 & 1.32 & 0.018
\\ \hline
unquenched & 0.187 & 1.30 & 0.016
\\ \hline
\end{tabular}

\vspace{0.20cm}

\begin{tabular}{|c|c|c|c|} \hline

$\stackrel{\mbox{V = 120 fm}^4}{\mbox{m = 0.65 }\Lambda_{QCD}}$
& $|\mathbf{\langle} \overline{\psi} \psi \mathbf{\rangle}|^{1/3}$ [GeV] 
& $M_N$ [GeV] & $Z$ [GeV$^6$]
\\ \hline
quenched & 0.204 & 1.27 & 0.018
\\ \hline
unquenched & 0.199 & 1.10 & 0.012
\\ \hline
\end{tabular}

\caption{Results of quenched and unquenched simulations for the quark 
condensate $\mathbf{\langle} \overline{\psi} \psi \mathbf{\rangle}^{1/3}$,
the neutron mass $M$ and the constant $Z$. The parameter $\Lambda_{QCD}$ is 
taken at the value $\Lambda_{QCD} = 220$ MeV.}

\label{tab:tab1}

\end{center}

\end{table}

\begin{table}

\begin{center}

\begin{tabular}{|c|c|c|c|} \hline

$\stackrel{\mbox{V = 56 fm}^4}{\mbox{m = 1.00 }\Lambda_{QCD}}$
& $\langle Q^2\rangle$ & $|d \widetilde{\Sigma}_\tau / dQ|$ [GeV$^8$] 
& $D / \overline{\theta}$ [e $\cdot$ fm] 
\\ \hline
quenched & 56 & 0.015 & 0.35 $\pm 0.06$
\\ \hline
unquenched & 22 & 0.025 & 0.32 $\pm 0.04$
\\ \hline
\end{tabular}

\vspace{0.20cm}

\begin{tabular}{|c|c|c|c|} \hline

$\stackrel{\mbox{V = 90 fm}^4}{\mbox{m = 0.75 }\Lambda_{QCD}}$
& $\langle Q^2\rangle$ & $|d \widetilde{\Sigma}_\tau / dQ|$ [GeV$^8$] 
& $D / \overline{\theta}$ [e $\cdot$ fm] 
\\ \hline
quenched & 90 & 0.043 & 0.79 $\pm 0.07$ 
\\ \hline
unquenched & 32 & 0.039 & 0.27 $\pm 0.07$
\\ \hline
\end{tabular}

\vspace{0.20cm}

\begin{tabular}{|c|c|c|c|} \hline

$\stackrel{\mbox{V = 120 fm}^4}{\mbox{m = 0.65 }\Lambda_{QCD}}$

& $\langle Q^2\rangle$ & $|d \widetilde{\Sigma}_\tau / dQ|$ [GeV$^8$] 
& $D / \overline{\theta}$ [e $\cdot$ fm] 
\\ \hline
quenched & 120 & 0.064 & 1.04 $\pm 0.06$
\\ \hline
unquenched & 40 & 0.046 & 0.17  $\pm 0.04$
\\ \hline
\end{tabular}

\caption{Results of quenched and unquenched simulations for the topological 
susceptibility $\langle Q^2 \rangle$, the quantity  $d \widetilde{\Sigma}_\tau 
/ dQ$ and the neutron EDM divided by the angle $\overline{\theta}$.
The statistical error in the last column is dominated by the uncertainty on the 
quantity $d \widetilde{\Sigma}_\tau / dQ$.}

\label{tab:tab2}

\end{center}

\end{table}

The quenched and unquenched results obtained for the EDM are also shown in 
Fig.~\ref{fig:edm} in order to better appreciate the quark mass dependence. 
The unquenched results appear to be consistent with a linear dependence on 
$m$ with a zero intercept, as expected from QCD and from the arguments leading 
to Eq.~(\ref{Dfull_chiral}). At the same time the quenched results exhibit a 
sharp increase at low quark masses, consistent with a divergence of the form 
given by Eq.~(\ref{Dq_chiral}) with $N_f = 2$. Clearly, a quenched calculation 
of the neutron EDM  becomes completely unreliable for $m_q \lesssim \Lambda_{QCD}$. 

In the quenched approximation, the divergence appearing in the chiral limit 
makes it impossible to perform the extrapolation toward the physical quark mass
value. On the other hand, such an extrapolation is possible in the case 
of unquenched results. Assuming a linear mass dependence, the IILM prediction 
for the neutron EDM corresponding to a light quark mass between 4 and 10 MeV 
is:
\be
|{\bf d}_n|=(6 \div 14) \times 10^{-3}~\ta~(\textrm{e}~\cdot~\textrm{fm}) ~ .
\ee

\section{Conclusions \label{sec:conclusions}}

We have used the Instanton Liquid Model to study the role played by the 
fermionic determinant in the evaluation of the neutron EDM, which is an 
observable sensitive to the topological structure of the vacuum. We have 
analyzed the chiral behavior of such a quantity (up to chiral logs) both 
in the quenched and unquenched cases. We have found that, when the fermionic 
determinant is included, the neutron EDM is expected to vanish linearly with 
the quark mass, whereas in the quenched approximation it should diverge as 
$1 / m^{N_f}$ in the chiral limit.

We have performed several model simulations and found that quenched and 
unquenched calculations give comparable results for the neutron EDM at large 
quark masses ($\simeq \Lambda_{QCD}$), whereas they strongly differ at lower 
quark masses. At the lowest value of the quark mass used in our simulations 
($ m \simeq 130$ MeV) the quenched result is a factor of $\simeq 4$ larger 
than the unquenched one. 

We have obtained the ILM prediction for the neutron EDM by extrapolating the 
unquenched IILM result to the physical value of the quark mass. The ILM result 
is roughly a factor $2 \div 4$ larger than existing model estimates \cite{Baluni,Crewther}.

Our main conclusion is that quenched and unquenched lattice QCD simulations 
of the neutron EDM as well as of other observables governed by topology might 
show up similar important differences in the quark mass dependence, near the 
chiral limit. In particular, our semi-classical analysis suggests that a 
quenched lattice calculation of the neutron EDM could be affected by a 
topology-driven divergence, which would make it impossible to perform the 
extrapolation to the physical value of the quark mass.

We insist on the fact that the qualitative predictions in Eq.~(\ref{Dfull_chiral}) 
and (\ref{Dq_chiral}) do not depend on the particular values of the model 
parameters which define the ILM. They rely only on a semi-classical description 
of the quantum mixing in the $\theta$-vacuum, in terms of isolated tunneling 
events. Hence, the observation of a divergence in a quenched lattice 
calculation of the neutron EDM, would represent a clean, parameter-free 
signature of instanton-induced dynamics in QCD.

\end{document}